\documentclass[showpacs,preprintnumbers,preprint,aps,amsmath,amssymb]{revtex4-1}
\usepackage{color}
\usepackage{axodraw}
\usepackage{graphics}
\usepackage{amsmath,amssymb,epsfig}
\newcommand{\mathsym}[1]{{}}

\newcommand{\ba}{\begin{array}}
\newcommand{\ea}{\end{array}}
\newcommand{\be}{\begin{equation}}
\newcommand{\ee}{\end{equation}}
\newcommand{\beqa}{\begin{eqnarray}}
\newcommand{\eeqa}{\end{eqnarray}}
\def\bt{\begin{table}}
\def\et{\end{table}}
\def\bc{\begin{center}}
\def\ec{\end{center}}
\def\bi{\begin{itemize}}
\def\ei{\end{itemize}}
\def\bea{\begin{eqnarray}}
\def\eea{\end{eqnarray}}
\def\beas{\begin{eqnarray*}}
\def\eeas{\end{eqnarray*}}

\def\tbar{$t\bar{t}~$}
\def\ubar{$u\bar{u}~$}
\def\dbar{$d\bar{d}~$}
\def\sbar{$s\bar{s}~$}
\def\qbar{$q\bar{q}~$}
\def\cbar{$c\bar{c}$}
\def\bbar{$b\bar{b}~$}
\def\pbar{$p\bar{p}~$}
\def\St{$\Sigma_3$}
\def\Sf{$\Sigma_4$}
\def\fba{A_{FB}}
\def\321{$SU(3)_c\times SU(2)_L\times U(1)_Y$}
\def\10{$SO(10)$}
\def\tbar{$t\bar{t}$ }
\newcommand{\phibar}  {\bar{\phi}}
\newcommand{\chibar}  {\bar{\chi}}
\newcommand{\sgm}  {\Sigma}
\newcommand{\sgmbar}  {\bar{\Sigma}}
\newcommand{\mubar}  {\bar{\mu}}
\newcommand{\nubar}  {\bar{\nu}}

\begin{document}
\vspace*{1cm}
\title{Forward-backward asymmetry in top quark production from light colored scalars in \10 model }
\bigskip
\author{Ketan M. Patel\footnote{kmpatel@prl.res.in} and Pankaj Sharma\footnote{pankajs@prl.res.in}}
\affiliation{Physical Research Laboratory, Navarangpura, Ahmedabad 380 009, India \vskip 1.0 truecm}

\begin{abstract}
\vskip 0.5 truecm 
The forward-backward asymmetry in top pair production at Tevatron has been reconfirmed by the CDF 
collaboration with 5.3 fb$^{-1}$ of accumulated data. These measurements also report that the 
asymmetry is the largest in regions of high invariant mass $M_{t\bar{t}}$ and rapidity difference
$|\Delta Y|$. 
We consider light colored sextet scalars appearing in a particular non-supersymmetric \10 grand unification model within 
the $\overline{126}$ scalar representation. These scalar states have masses in the range of 
$300~ \text{GeV}-2 ~\text{TeV}$ consistent with the requirements of gauge coupling unification 
and bounds on the proton lifetime. The cross section and the total asymmetry can be simultaneously explained 
with the contributions of these scalars within 1$\sigma$. We find that the simultaneous fitting of 
the cross section, the total asymmetry and the asymmetries in different rapidity and $M_{t\bar{t}}$ bins gives 
only a marginal improvement over the SM contribution. We also study various production mechanisms of these colored sextet scalars at the LHC.
\end{abstract}

\maketitle

\section{Introduction}

The CDF collaboration had measured the forward-backward asymmetry, $\fba$, in \tbar pair production 
in the \tbar rest frame in 2008 at Tevatron with 3.2 fb$^{-1}$ of collected data, as
 \cite{Aaltonen:2008hc}
\begin{equation}\label{afb}
 A_{FB}\equiv\frac{N(\cos\theta>0)-N(\cos\theta<0)}{N(\cos\theta>0)+N(\cos\theta<0)}
=0.193\pm 0.065(\text{stat})\pm 0.024(\text{syst})
\end{equation}
where $\theta$ is the scattering angle of top quark in the \tbar rest frame.  
This result had been confirmed by D0 collaboration based on 0.9 fb$^{-1}$ of integrated 
luminosity \cite{:2007qb}. They reported $\fba=0.19\pm 0.09(\text{stat})\pm0.02(\text{syst})$ and  
$\fba=0.12\pm 0.08(\text{stat})\pm0.01(\text{syst})$ for exclusive 4-jet and inclusive 4-jet events 
respectively consistent with CDF results. These measurements have attracted a lot of attention due to more than 2$\sigma$
deviation from the Standard Model (SM) predicted value of $\fba^{SM} = 0.058\pm0.009$
 \cite{smvalue}. In the SM, the $\fba$ identically vanishes at leading order (LO). However, at next
to
leading order (NLO) in QCD, it can arise from (a) the interference between tree level SM amplitude
and the box diagram,  (b) radiative corrections to \qbar annihilation and (c) interference between 
different amplitudes contributing to gluon-quark scattering. Several independent New
Physics (NP) scenarios have been advanced \cite{npfba1,tait, jung, dorsner1, dorsner2, arhrib} to
explain this
discrepancy.

Recently, CDF have presented new results using 5.3 fb$^{-1}$ of data sets in which $\fba$ is
reported to be $0.158\pm0.074~(\text{stat+syst})$. Along with this new value of $\fba$, they find
interesting dependences of $\fba$ on the invariant mass of the \tbar pair and their rapidity dependence. The
asymmetry is more prominent in the large invariant mass region of $M_{t\bar{t}}>450$ GeV with more
than
3$\sigma$ deviation and in the large rapidity difference $|\Delta Y|>1$ region with around 2$\sigma$
deviation from the SM predicted value \cite{Aaltonen:2011kc}. On the other hand, some other observables
related to \tbar pair production at Tevatron show good agreement with the SM predicated values. The
measured parton level \tbar-pair production cross section $\sigma_{t\bar{t}}^{exp}=7.70\pm0.52$
 \cite{Aaltonen:2010ic} agrees with the SM predicted value of
$\sigma_{t\bar{t}}^{SM}=7.45^{+0.72}_{-0.63}$ calculated with MCFM \cite{mcfm}. Similarly the
experimentally
measured invariant mass distribution is also consistent with the prediction of the
SM at NLO \cite{Aaltonen:2009iz}. Hence, while discussing the new physics scenarios to explain the
new results, namely the mass and rapidity dependence of $\fba$, they must not introduce large
corrections to either the total \tbar cross section $\sigma_{t\bar{t}}$ or the invariant mass
 distribution $M_{t\bar{t}}$. Some recent attempts \cite{npfba2, Grinstein} have been made in order
to explain these new observables along with the updated measurements of $\fba$ and
$\sigma_{t\bar{t}}$.

The standard model gauge structure (\321) allows a finite number of different representations of
scalar particles which can couple an up or down quark to the top quark. The possible cases include a
set of colored octet, singlet, triplet and sextet scalars each for $u\bar{u} \rightarrow t\bar{t}$
and
$d\bar{d} \rightarrow t\bar{t}$ processes.
\be \label{scalarrep}
\ba{cc}
(8,2,\frac{1}{2}),~~(1,2,\frac{1}{2}),~~(\bar{3},3,\frac{1}{3}),~~(6,3,\frac{1}{3}
)&~~~~~\text{for}~u\bar{u}/d\bar{d} \rightarrow
t\bar{t}\\
(\bar{3},1,\frac{4}{3}),~~(6,1,\frac{4}{3})&~~~~~\text{for}~u\bar{u} \rightarrow
t\bar{t}~\text{only}\\
(\bar{3},1,\frac{1}{3}),~~(6,1,\frac{1}{3})&~~~~~\text{for}~d\bar{d} \rightarrow
t\bar{t}~\text{only}\\
\ea \ee 
The effects of these scalars on $\fba$ have been studied in a model independent way in several
papers.
For example, the results of Shu {\it et al} \cite{tait} show that colored sextet and triplet
scalars are able to explain the anomaly while the analysis of Jung {\it et al} \cite{jung} favours
the singlets and sextets. On the other hand, the results of Arhrib {\it et al} \cite{arhrib} show that sextet
diquarks could not fit the $\fba$ and cross section simultaneously within $1\sigma$. However all
these analyses were based on old observations of $\fba$. According to the new CDF data, the central
value of $\fba$ has significantly come down. Also the distributional preferences of $\fba$ in
invariant mass and rapidity have been reported. So it is an interesting exercise to study
these scalars in the light of new observations. In the present study we investigate the light
colored sextet scalars appearing in a well motivated \10 grand unified theories as a possible
explanation of $\fba$ as well as new observables simultaneously based on current data.

Colored scalar fields naturally emerge in a well motivated class of grand unified theories. For
example, the representations $(8,2,\frac{1}{2})$ and $(\bar{3},1,\frac{4}{3})$ reside in a
45-dimensional Higgs field of $SU(5)$. It is interesting to note that in any simple renormalizable
version of non-supersymmetric $SU(5)$, the 45 Higgs together with 5-dimensional Higgs is necessarily
required to generate viable masses of charged fermions \cite{gj}. It has been shown through detailed
studies in reference \cite{dorsner1, dorsner2} that both the scalar states $(8,2,\frac{1}{2})$ and
$(\bar{3},1,\frac{4}{3})$ can have masses in the range of 300 GeV - 1 TeV consistent with
the requirements of gauge coupling unification and bound on the proton lifetime. In addition, the
contribution of colored triplet scalar to the production of $t\bar{t}$ at the Tevatron can enhance
the forward-backward asymmetry and account for the experimental result without spoiling the successful
standard model prediction for the total cross section. We investigate a similar possibilities in more
predictive and attractive class of grand unified theories based on the \10 gauge group. The
remarkable feature of \10 is that its 16-dimensional irreducible spinor representation
accommodates a complete family of fermions, including the right-handed neutrino. This complete
unification of quarks and leptons opens up the possibly of connections between the charge fermions
and the neutrino sector. Furthermore, \10 has the left-right symmetry group $SU(2)_L\times SU(2)_R$ 
as a subgroup, making the implementation of the both the type-I and the type-II seesaw mechanisms
very 
natural in these theories. The 45 dimensional scalar representation of $SU(5)$
resides in both the $\overline{126}$ and the $120$ dimensional scalar representations
of \10 which can couple with ordinary fermions through the Yukawa interactions. The
$\overline{126}$ Higgs field plays an important role in gauge symmetry breaking
 \cite{bertolini} as well as it is essential for viable fermion masses \cite{jp}.
However the color triplets of $\overline{126}$ couple with fermions through
symmetric leptoquark couplings  and can induce the rapid proton decay if assumed
light. On the other hand, the $120$ Higgs field has antisymmetric coupling with
fermions but it is not required if one sticks to the minimal Higgs content of
the theory. We show in this work that the  $\overline{126}$ Higgs has diquark colored sextets and
octets at TeV scale consistent with gauge coupling
unification and proton decay bounds and study the role of the sextet scalars
as the possible candidates to explain the anomaly in \tbar production observables.

This paper is organized as follows. In the next section, we study the scalar spectrum of a
particular \10 model and discuss the constraints coming from gauge coupling unification and proton
decay. In section III, we study the role of light colored sextet scalars on the \tbar pair production 
observables. In section IV, we will study the signatures of these scalars at both 7 TeV and 14 TeV 
LHC. Finally, we summarize our results in section V.

\section{Light Colored Scalars in \10 model}

We consider non-supersymmetric \10 as a basic framework of our model. It has been
pointed out in recent studies \cite{bertolini} that an adjoint 45-dimensional scalar
representation ($\chi$) of \10 together with one $\overline{16}$ or $\overline{126}~(\sgm)$ Higgs
can govern
the entire breaking of \10 gauge symmetry down to the SM. If one sticks to the renormalizable
version of the seesaw mechanisms then the representation $\overline{126}$ is indispensable, since it
breaks the $SU(2)_R$ group and gives neutrino masses through seesaw mechanisms. In addition, one
needs $10$ dimensional Higgs ($\phi$) to obtain a realistic fermion mass spectrum \cite{jp}. We have
given the decompositions and full SM spectrum of these scalar fields in Table \ref{feilds} in
Appendix. Note that $\chi$ contains two SM singlets ($\chi_3, \chi_8$) and $\Sigma$ contains one ($\Sigma_9$) SM
singlet that acquire vevs at GUT scale and break \10 to the SM group. The $10$ and
$\overline{126}$ Higgs contain the SM doublets ($\phi_2, \bar{\phi}_2$) and ($\Sigma_2,
 \bar{\Sigma}_2$) respectively which can mix through a renormalizable term $\chi_{ij} \chi_{kl}
\Sigma_{ijklm}
\phi_m$ in the scalar potential. For consistent fermion mass spectrum, one has to keep (at least)
one linear combination of these doublets light upto the electroweak scale, which
plays the role of SM Higgs doublet and triggers the electroweak symmetry breaking. This requires a
fine tuning in the parameters of the Higgs potential. Assuming such fine tuning in parameters, a
detailed numerical analysis has been carried out recently for viable fermion mass spectrum in this
model in Ref. \cite{jp}. It has been shown that such model can provide very predictive
structure of fermion masses if a global $U(1)_{PQ}$ (Peccei-Quinn) symmetry is imposed and can
produce realistic fermion mass spectrum which is in excellent agreement with the present data
extrapolated at the GUT scale.

We consider a non-supersymmetric \10 framework with the minimal Higgs fields $10+45+\overline{126}$ in
our attempt to explain the forward-backward asymmetry in $t\bar{t}$ production at Tevatron.
Following previous studies, we assume that only one linear combination of the weak doublets of 10
and $\overline{126}$ remains light and becomes the SM Higgs. Further, we also need to
assume that the scalar submultiplets which can potentially contribute to the asymmetry in the production
of top quarks also remain light in the range of 300 GeV - 2 TeV. Among the possible options allowed
by the SM gauge symmetry shown in Eq. \ref{scalarrep}, the $\overline{126}$ contains three sextets
$\sgm_3 (6,1,\frac{4}{3})$, $\sgm_4 (6,1,\frac{1}{3})$, $\sgm_{12} (\bar{6},3,-\frac{1}{3})$ and
a pair of a octets $\sgm_{15} (8,2,\frac{1}{2})$, $\sgmbar_{15} (8,2,-\frac{1}{2})$. These fields
couple to the $16$-plet matter through Yukawa
interactions. Furthermore, all these fields are diquark (have only quark-quark coupling) in nature
and do not mediate the proton decay. Some other components of the $\overline{126}$ scalar (like $\sgm_1,
\sgm_7$ and $\sgm_{13}$) also have the correct quantum numbers to influence $t\bar{t}$ production
but they have leptoquark coupling which induce rapid proton decay if assumed light. If no artificial
suppression via Yukawa couplings is arranged their masses should not be below than $10^{12}$ GeV due
to proton decay constraints.

The interaction of $\overline{126}$ Higgs field to the 16-dimensional matter fields $\psi$ can be
written in its most general form as \cite{Mohsakita}
\be \label{yukawa}
-{\cal L}_Y = \dfrac{1}{5!}F_{ij} \psi^T_{i}{\cal B}C^{-1}\Gamma_p\Gamma_q\Gamma_r\Gamma_s\Gamma_t
\psi_{j} \Sigma_{pqrst}
\ee
where the indices $i,j$ denote family indices, $p,q,..=1,..,10$ are \10 indices, $C$ is the
Dirac charge conjugation matrix and ${\cal B}=\Gamma_1\Gamma_3\Gamma_5\Gamma_7\Gamma_9$ is the
equivalent of the charge conjugation matrix for \10 that ensures the invariance under \10.
$\Gamma_i$'s are representations of the Clifford algebra associated with the Lie algebra of the 
\10 group and are given in \cite{Mohsakita, Wilczek}. $F$ is the Yukawa coupling matrix and 
it is symmetric by its \10 properties. After the decomposition \cite{decomp} of Eq. \ref{yukawa}, 
the couplings of $\sgm_3$, $\sgm_4$, $\sgm_{12}$ and $\sgm_{15}$ to matter can be written as
\beqa \label{intterm}
-{\cal L}_Y &\ni &- 2 F_{ij}~u_{ai}^{CT} C^{-1} u_{bj}^C {\sgm_3}_{ab},\nonumber \\ 
	    & &\sqrt{2} F_{ij}~u_{ai}^{CT} C^{-1} d_{bj}^C {\sgm_4}_{ab},\nonumber \\
	    & &- 2F_{ij}~Q_{ai}^{T} C^{-1} \varepsilon{\sgm_{12}}_{ab} Q_{bj},\nonumber \\
	    & &-2F_{ij}~(u_{ai}^{T} (T^A)_{ab} C^{-1} u_{bj}^C \sgm_{15}^{0A}+d_{ai}^{T}
(T^A)_{ab} C^{-1} u_{bj}^C \sgm_{15}^{+A}) \nonumber \\
& &\sqrt{2} F_{ij}~(d_{ai}^{T} (T^A)_{ab} C^{-1} d_{bj}^C \sgmbar_{15}^{0A}+u_{ai}^{T}
(T^A)_{ab} C^{-1} d_{bj}^C \sgmbar_{15}^{-A})
\eeqa
where $T^A=\frac{1}{2}\lambda^A$ and $\lambda^A~(A=1,..,8)$ are the Gell-Mann matrices of $SU(3)$.
$a,b,c$ are color indices. Clearly, all these fields have the right couplings to influence the
asymmetry we are interested in. However, in order to be relevant for asymmetry at Tevatron, all
these fields or at least one of
them must be sufficiently light. On the other hand, such light states will contribute to the running
of gauge couplings and hence viability of their being light is constrained by the unification of
gauge couplings and present bound on the proton lifetime. We thus show that it is possible to
achieve
light colored scalars with successful gauge coupling unification in a consistent way in our model.

In the absence of any new particle thresholds between the weak and GUT scales, the running of gauge
couplings at one-loop level is given by 
\be \label{running}
\alpha_{GUT}^{-1}=\alpha_{i}^{-1}(M_Z)-\dfrac{b_i}{2\pi}\ln\left( \dfrac{M_{GUT}}{M_Z}\right) 
\ee
where $\alpha_{GUT}$ represents the
gauge coupling at the unification scale $M_{GUT}$. $b_i$'s are the appropriate one-loop $\beta$ function
coefficients \cite{chengli} and $i=1,2,3$ stands for $U(1)_{Y}$, $SU(2)_{L}$ and $SU(3)_{c}$
respectively. Their values for the SM with one light Higgs doublet are $b_1=\frac{41}{10}$,
$b_2=-\frac{19}{6}$ and $b_3=-7$. It is easy to check that these values for $b_i$ do not unify the
gauge couplings since SM does not predict gauge coupling unification in the first place. The presence
of new particles between weak and GUT scale can change the running and it can be easily
incorporated by replacing $b_i$ in Eq. \ref{running} with effective one-loop coefficients $B_i$
defined by \cite{giveon}
\be \label{beta}
B_i=b_i+\sum_I b_{iI} \dfrac{\ln(M_{GUT}/M_I)}{\ln(M_{GUT}/M_Z)}
\ee
where $b_{iI}$ is the one-loop coefficient of the additional particle $I$ of mass $M_I$ lying between $M_Z$ and
$M_{GUT}$. Following Giveon {\it et al.} \cite{giveon}, Eq. \ref{running} with contributions from
Eq. \ref{beta} can provide successful gauge coupling unification at one loop level if they satisfy
following two conditions:
\beqa \label{conditions}
\dfrac{B_{23}}{B_{12}} &\equiv& \dfrac{B_2-B_3}{B_1-B_2} = \dfrac{5}{8}
\dfrac{\sin^2\theta_W-\alpha/\alpha_s}{3/8-\sin^2\theta_W} = 0.716\pm0.005 \\
B_{12} &\equiv& B_1-B_2 = \dfrac{16\pi}{5 \alpha}
\dfrac{3/8-\sin^2\theta_W}{\ln(M_{GUT}/M_Z)} = \dfrac{184.9\pm0.2}{\ln(M_{GUT}/M_Z)}\eeqa
We use the present experimental measurements of the SM parameters \cite{pdg} to derive the
above numbers. In any given model, $B_{ij}$ depend only on the particle content and associated mass
spectrum and conditions (\ref{conditions}) allow us to constrain the mass spectrum of particles
that leads to an exact unification at GUT scale. We give a list of the different submultiplets of
$10$, $\overline{126}$ and $45$ scalar representations and the corresponding contributions to
coefficients $B_{ij}$ in Table(\ref{feilds}) in the Appendix.

In order to present a consistent analysis, we now discuss the constraints coming from proton decay. In
nonsupersymmetric GUTs, this process is mediated by baryon number violating gauge interactions
which induce a set of effective dimension six operators at low energies that conserve $B-L$. In
the \10
scenario we consider here, such gauge bosons are integrated out at the GUT scale
($m_{X,Y}=M_{GUT}$) and therefore proton decay constrains $M_{GUT}$ from below. The most stringent
bounds coming from the latest experimental limit on partial decay lifetime of proton $\tau_p$
$(p\rightarrow\pi^0e^+) > 8.2 \times 10^{33}$ years \cite{pdecay} implies
\be \label{pd}
M_{GUT}\approx (m_p^5 \alpha_{GUT}^2 \tau_p)^{\frac{1}{4}} \gtrsim
~2.3\times10^{16}~\sqrt{\alpha_{GUT}} ~~ {\rm GeV,}
\ee
where $m_p=0.938$ GeV is the proton mass. Some of the submultiplets of $10$ and $\overline{126}$
Higgs are leptoquark scalars (for example, $\phi_1, \sgm_1, \sgm_{6}$ and etc.) and are 
associated with $d=6$ proton decay operators. We suppress their contribution to proton decay by
making them super heavy $\sim M_{GUT}$ as we will explain in next paragraph.

As mentioned earlier in this section, it is necessary that at least one of the submultiplets 
($\sgm_3, \sgm_4, \sgm_{12}$ and $\sgm_{15}$) of $\overline{126}$ remains light in order to explain
the forward-backward asymmetry in top quark production. Typically in theories with two or more
widely different mass scales, if a submultiplet of a full Higgs multiplet acquires a vev $\simeq M$,
the members of that multiplet acquire a mass $\sim M$ \cite{mg}. Any scenario which differs from
this would require some fine tunings in the parameters of the scalar potential. To check the viability
of such fine tunings, a complete detailed analysis of scalar potential minimization and its
diagonalization is required. However this is beyond the scope of present work and we assume
that such fine tuning is possible in our case. In order to avoid further unnecessary fine tunings,
we assume that the remaining submultiplets of the scalar fields $10, 45, \overline{126}$ are super heavy
and have natural masses of order $M_{GUT}$. In other words, we assume that only those submultiplets
of scalar fields remain light and have masses $M_I$ in between the weak scale and the GUT scale
and may potentially contribute to the forward-backward asymmetry in $t\bar{t}$ production.

\begin{figure}[ht]
 \centering
 \includegraphics[width=4.25in]{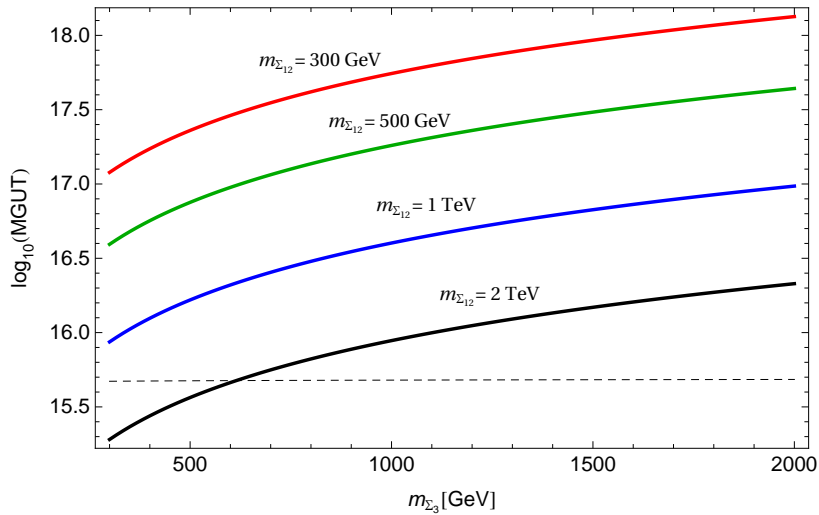}
 \caption{The maximum value of $M_{GUT}$ obtained for different values of $m_{\sgm_{12}}$ by
assuming \St ~light. The dashed line stands for the lower bound on $M_{GUT}$ due to the proton
lifetime. Viable gauge coupling unification is achieved in the region between $m_{\sgm_{12}}$=300
GeV and the dashed line.}
 \label{fig1}
\end{figure}

\begin{figure}[ht]
 \centering
 \includegraphics[width=4.25in]{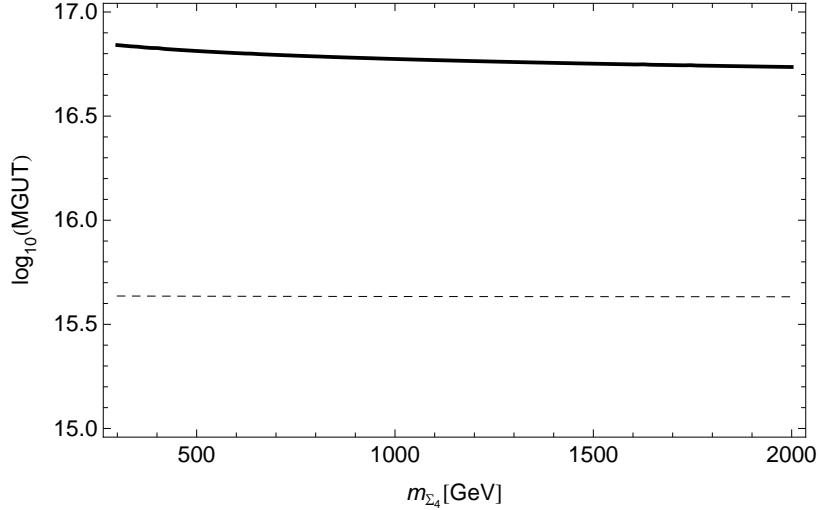}
 \caption{The maximum value of $M_{GUT}$ obtained by assuming \Sf~ light. The dashed line stands for
the lower bound on $M_{GUT}$ due to the proton lifetime. Viable gauge coupling unification is
achieved in the region between two lines.}
 \label{fig2}
\end{figure}

With all these considerations, we now check the compatibility of light colored scalar states of our
interest with the unification of the gauge couplings and constraints on GUT scale coming from
proton decay bounds. Following the strategy of \cite{dorsner1}, we determine an upper bound on GUT
scale at the one loop level assuming that any one of $\sgm_3, \sgm_4, \sgm_{12}$ and $\sgm_{15}$ is
responsible for asymmetry and is accordingly in the mass range of 300 GeV - 2 TeV. For this, we
numerically maximize $M_{GUT}$ while imposing the condition that the solution satisfies
Eq. \ref{conditions}. The
additional constraints we put on the solution are $300~{\rm GeV} \leq m_{\sgm_3}, m_{\sgm_4},
m_{\sgm_{12}}, m_{\sgm_{15}}= m_{\sgmbar_{15}} \leq M_{GUT}$ and $M_{GUT}\leq M_{Planck} =
10^{19}~{\rm GeV}$. The
results of our numerical analysis are shown in Fig. \ref{fig1} and Fig. \ref{fig2}. We get viable
gauge coupling unification consistent with proton decay limits in two different scenarios.


\noindent(A) We get successful gauge coupling unification for the scalar diquark $\sgm_3$ having 
mass in the range of $300~{\rm GeV}$ to $2~{\rm TeV}$ as shown in Fig. \ref{fig1}. The light
$\sgm_3$ also requires light $\sgm_{12} (\bar{6},3,-\frac{1}{3})$ and there is a clear correlation
between their masses. There exists an upper bound on $m_{\sgm_{12}}$ for a given value of
$m_{\sgm_{3}}$. For example, when $m_{\sgm_{3}} = 600$ GeV we have $m_{\sgm_{12}} \leq 2$ TeV. The
other two scalar states remain heavy, namely, $\sgm_{15}\sim 10^{9}-10^{12}$ GeV and $\sgm_{4}\sim
M_{GUT}$.

\noindent(B) Unification of gauge coupling is also achieved with light sextet diquark
state $\sgm_{4} (6,1,\frac{1}{3})$ as shown in Fig. \ref{fig2}. Unlike light $\sgm_3$ in the
previous case it
does not require any other light submultiplet at TeV scale. The maximum value of the GUT scale does not change
appreciably with $m_{\sgm_{4}}$ and stays well above the present proton decay limits
shown by the dashed line in Fig. \ref{fig2}. Successful unification in this case requires only one
state at intermediate scale $m_{\sgm_{12}} \sim 10^{8}$ GeV and $\sgm_{3}, \sgm_{15}$ and remain
superheavy ($\sim M_{GUT}$).

From the results of the detailed analysis carried out in this section, we conclude that either
$\sgm_{3}$ or $\sgm_{4}$ can remain light and influence the forward-backward asymmetry in
$t\bar{t}$ production through the processes $u\bar{u} \rightarrow t\bar{t}$ and $d\bar{d}
\rightarrow t\bar{t}$ respectively. We do not get viable gauge coupling unification with a light
colored octet state $\Sigma_{15}$ and hence this case will be ignored in further analysis. Note that we
have presented consistent unification analysis at the one loop level. The allowed masses of \St ~and
\Sf~ would change slightly if one considers two-loop effects in the running of gauge couplings.
However they would still remain within the TeV range.

\section{Colored sextets and Forward-Backward Asymmetry of top quarks }

\begin{figure}[htb]
\begin{center}
\begin{picture}(800,130)(0,0)

\ArrowLine(310,105)(350,80)
\ArrowLine(390,105)(350,80)
\Vertex(350,80){2}
\DashLine(350,50)(350,80){5}
\Vertex(350,50){2}
\ArrowLine(350,50)(390,20)
\ArrowLine(350,50)(310,20)
\put(295,90){$d,s,b$}
\put(295,30){$\bar{d},\bar{s},\bar{b}$}
\put(390,35){$t$}
\put(390,85){$\bar{t}$}
\put(360,60){\Sf}
\put(350,00){$(b)$}


\ArrowLine(50,105)(90,80)
\ArrowLine(130,105)(90,80)
\Vertex(90,80){2}
\DashLine(90,50)(90,80){5}
\Vertex(90,50){2}
\ArrowLine(90,50)(130,20)
\ArrowLine(90,50)(50,20)
\put(30,90){$u,c$}
\put(35,30){$\bar{u},\bar{c}$}
\put(130,35){$t$}
\put(130,85){$\bar{t}$}
\put(100,60){\St}
\put(90,00){$(a)$}
\end{picture}
\caption{ Contributions from light sextet scalars to the 
\tbar production at the Tevatron.}\label{feyngraph}
\end{center}
\end{figure}
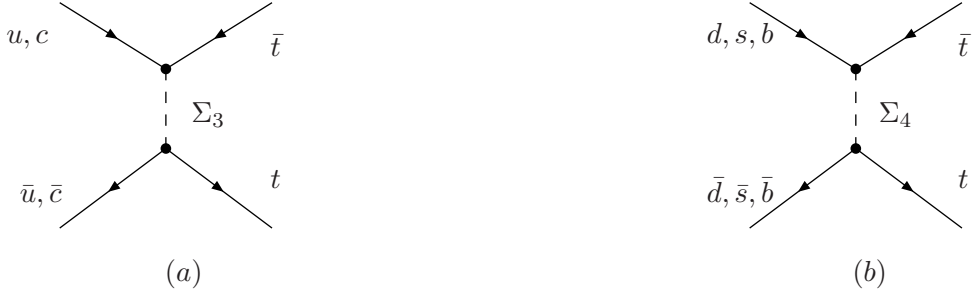

The light colored sextet scalars \St ~and \Sf ~contribute to \tbar pair production through u-channel
exchange as shown in Fig. \ref{feyngraph}. \St ~interferes with the SM contributions for \ubar and
\cbar~ initial parton states while \Sf ~interferes with the SM contributions for \dbar, \sbar and 
\bbar initial parton states. The contributions of initial parton states \cbar, \sbar and \bbar to
the overall process \pbar $\rightarrow$ \tbar will be suppressed due to their small parton
distribution functions (PDF). However we include all these contributions in our analysis.

%
%

Let us denote incoming quark momentum by $p_{q}$, incoming anti-quark 
momentum by $p_{\bar{q}}$, outgoing top momentum by $p_{t}$ and outgoing 
anti-top momentum by $p_{\bar{t}}$ with the following definitions :
\bea\label{momdef}
p_{q,\bar{q}}&=&\frac{\sqrt{\hat{s}}}{2}(1,0,0,\pm 1),\\
p_{t,\bar{t}}&=&[E_{t},\pm|\overrightarrow{p_{t}}|(\sin\theta,0,\cos\theta)],\\
	     &\equiv& E_t[1,\pm\beta_t(\sin\theta,0,\cos\theta)]
\eea 
where $E_t=\sqrt{\hat{s}}/2$, 
$\beta_t=|\overrightarrow{p_{t}|}/E_t\equiv\sqrt{1-4m_t^2/\hat{s}}$ 
and $\theta$ is the angle between momenta of the incoming quark and the outgoing 
top quark in the center of mass (cm) frame of the partons. Neglecting masses of 
all quarks except the top, the Mandelstam variables in parton cm 
are defined as follows :
\bea\label{Mandel}
\hat{s}&=&(p_q+p_{\bar{q}})^2=(p_t+p_{\bar{t}})^2=x_1x_2s,\\
\hat{t}&=&(p_q-p_t)^2=(p_{\bar{q}}-p_{\bar{t}})^2=m_t^2-\frac{\hat{s}}{2}(1-\beta_t\cos\theta),\\
\hat{u}&=&(p_q-p_{\bar{t}})^2=(p_{\bar{q}}-p_t)^2=m_t^2-\frac{\hat{s}}{2}(1+\beta_t\cos\theta)
\eea
where $s$ is the cm energy of proton and antiproton in laboratory frame, 
$x_1$ and $x_2$ are the fractions of momentum carried by the partons inside proton 
and antiproton respectively.

With these notations and conventions, the matrix amplitude squared (averaged and summed over 
initial and final color and spin indices respectively) for (\qbar$\rightarrow$ \tbar) can be written as follows :
\bea \label{Msq}
\overline{\sum}|\mathcal M_{total}|^2&=&\nonumber \frac{2g_s^4}{9}
\left[1+\frac{4m_t^2}{\hat{s}}\sin^2\theta+\cos^2\theta\right]\\\nonumber
&-&\frac{4g_s^2}{9}\frac{|{f^{u,d}_{13}}|^2}{\hat{u}-m_{\Sigma_{3,4}}^2}
\left[(1+\beta_t\cos\theta)^2+\frac{4m_t^2}{\hat{s}}\right]\\
&+&\frac{|{f^{u,d}_{13}}|^4}{12(\hat{u}-m^2_{\Sigma_{3,4}})^2}(1+\beta_t\cos\theta)^2,
\eea
where $f^{u}_{13}$ and $f^{d}_{13}$ are related with the original coupling $F_{13}$ of
Eq. \ref{intterm} by the following relation
\be \label{coupling}
f^{u}_{13}=\sqrt{2}f^{d}_{13}=2F_{13}.
\ee
%

For our numerical study, we have used the leading order PDF sets of CTEQ6L \cite{cteq6l} to
convolute with the partonic cross section to obtain hadronic cross section. We set our
renormalization and factorization scale to $\mu_R=\mu_F=m_t$. The top mass is taken to be
$m_t=172.5$ GeV at which we also evaluate strong coupling $\alpha_s=0.1085$. We use $K-$factor of
1.3 to rescale our LO results for $\sigma(t\bar{t})$ to match with NLO QCD prediction
 \cite{Kidonakis:2008mu}. 
%


We calculate the total cross section $\sigma$(\tbar), $\fba$ as defined in Eq. \ref{afb}, as well as
$\fba$ in $|\Delta Y|>1$,
$|\Delta Y|<1$, $M_{t\bar{t}}<450$ GeV and $M_{t\bar{t}}>450$ GeV, where $|\Delta Y|$ is the
difference of top and anti-top quark rapidities i.e., $|\Delta Y|=Y_t-Y_{\bar{t}}$ in \tbar rest
frame. The present experimentally measured values of all these observables, their values predicted
in the SM and corresponding contributions needed from NP are listed in Table \ref{input}.

\begin{table} [ht]
\begin{small}
\begin{math}
\begin{array}{|l|c|c|c|}
\hline
 \text{Observables} & \text{Experimentally} & \text{SM Contribution} & \text{Contribution needed
} \\
 \text{ } & \text{Measured Values} & \text{ } & \text{from NP} \\
\hline
\text{Cross section}	&7.70\pm0.52 		& 7.45^{+0.72}_{-0.63}			& -  \\
A_{FB}			&0.158\pm0.074 		& 0.058\pm0.009		&  0.1\pm0.083	 \\
A_{FB} (M_{t\bar{t}}>450 {\rm GeV})	&0.475\pm0.112 		& 0.088\pm0.0013		
& 0.387\pm0.1133 \\
A_{FB} (M_{t\bar{t}}<450 {\rm GeV})	&-0.116\pm0.153 		& 0.04\pm0.006
	
& -0.156\pm0.159 \\
A_{FB} (|\Delta Y|>1)	&0.611\pm0.256 		& 0.123\pm0.018  &0.488\pm0.274 \\
A_{FB} (|\Delta Y|<1)	&0.026\pm0.118		& 0.039\pm0.006  &-0.013\pm0.124 \\
\hline
\end{array}
\end{math}
\end{small}
\caption{The observables with their experimentally measured values, their values predicted
in the SM and corresponding contributions needed from NP. The contributions needed from NP are
obtained by subtracting the SM contributions from experimentally measured values.}
\label{input}
\end{table}

We perform a $\chi^2$ analysis to simultaneously fit all the observables  shown in
Table \ref{input}. For this, we define the following $\chi^2$ function
\be \label{chisq}
\chi^2=\sum_{i=1}^{6} \left(\frac{P_i-O_i}{\sigma_i}\right)^2 ~,\ee
where the sum  runs over all the six observable quantities. $P_i$'s are the theoretically 
calculated values of these quantities as a function of couplings and masses of scalars in our model
and $O_i$'s are the mean values of these
observables. $\sigma_i$'s denote $1\sigma$ errors in $O_i$. The $\chi^2$ is numerically minimized 
to obtain the best fit over all six observables. The more robust statistic to quantify the quality of fit is 
reduced-$\chi^2$ which is defined as the $\chi^2/\nu$ where $\nu$ is number of degrees of freedom (d.o.f.) in the analysis. 
For the SM, the value of total $\chi^2$ is 17.26 and the value of $\chi^2/\nu$ is $2.88$.

We now present a detailed numerical analysis for the contributions of \St ~ and \Sf~ separately. 
\subsection{Diquark $(6,1,\frac{4}{3})$}
\begin{figure}[h]
\begin{center}
\includegraphics[width=3.0in]{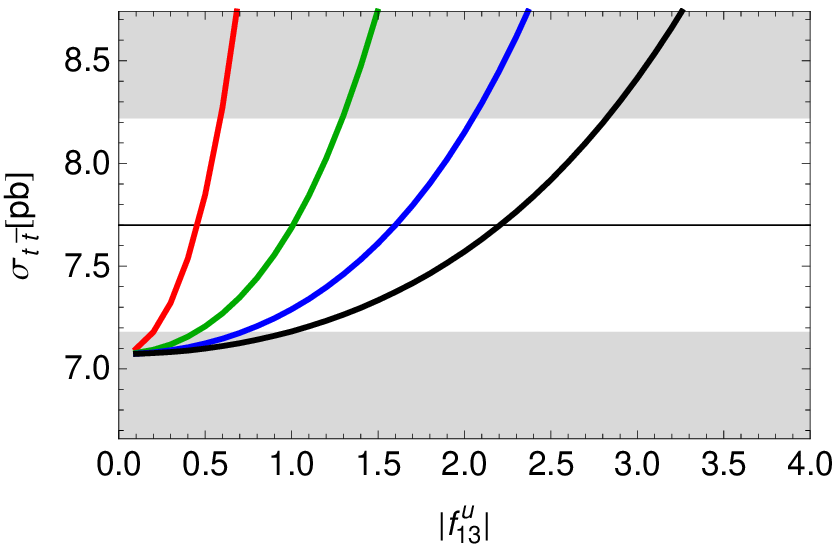}
\includegraphics[width=3.15in]{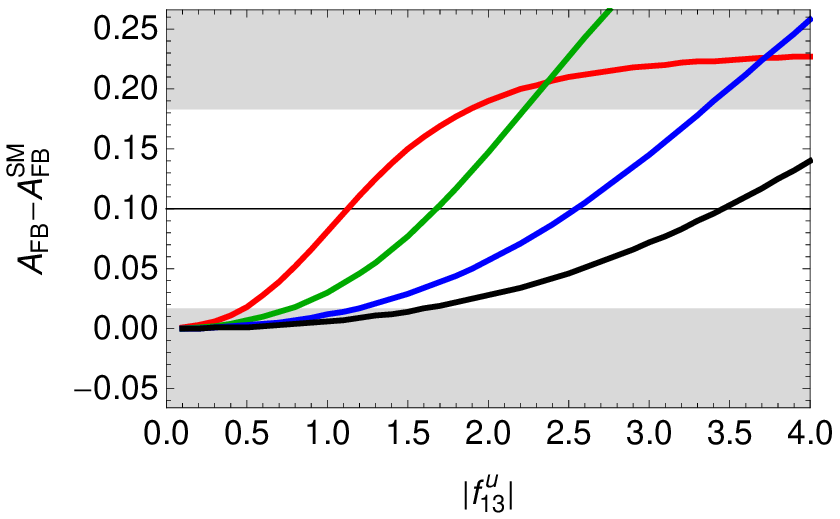} 
\caption{ The \tbar production cross section (left) and the forward backward asymmetry (right) 
as a function of the coupling $f^u_{13}$ for masses $m_{\Sigma_3}= $ 300 GeV (Red), 900 GeV (Green), 
1500 GeV (Blue) and 2100 GeV (Black). The unshaded and the shaded region correspond to $1\sigma$ 
and $2\sigma$ experimental bounds respectively.} 
\label{cs-afb-uu}
\end{center}
\end{figure}

In Fig. \ref{cs-afb-uu}, we plot the cross section and the forward backward asymmetry for \tbar
production at Tevatron as a function of the coupling $f^u_{13}$ for four different masses of 
the colored sextet scalar $\Sigma_3$. In showing the contribution from new physics, we subtract 
the SM contribution from the experimentally measured value of $\fba$. 

From Fig. \ref{cs-afb-uu}, we see that sextet \St ~of mass 300 GeV can barely satisfy both 
$\sigma^{t\bar{t}}$ and $\fba$ constraints for a very narrow range of coupling $f^u_{13}$ 
that too at 2$\sigma$. For large masses \St ~can satisfy both the constraints for large range of 
coupling $f^u_{13}$ within 1$\sigma$ of the experimental bound.

\begin{table} [h]
\begin{small}
\begin{math}
\begin{array}{|l|cc|cc|cc|cc|}
\hline
 &  \multicolumn{2}{c|} {m_{\Sigma_3}=300 \text{GeV}} &
\multicolumn{2}{c|} {m_{\Sigma_3}=900
\text{GeV}} & \multicolumn{2}{c|} {m_{\Sigma_3}=1.5 \text{TeV}} &
\multicolumn{2}{c|}
{m_{\Sigma_3}=2.1 \text{TeV}} \\
\hline
 \text{Observables} &  \text{Fit} & \text{pull} &
\text{Fit} &
\text{pull} & \text{Fit} & \text{pull} & \text{Fit} &
\text{pull} \\
\hline
\text{Cross section} & 8.0404 & 0.6546 & 8.2761 & 1.1078 &
8.3004 & 1.1546 & 8.3076 &
1.1685 \\
A_{FB} & 0.0237 &-0.919 & 0.0617 &-0.462 & 0.0690
&-0.3729 & 0.0715 &-0.3429 \\
A_{FB} (M_{t\bar{t}}>450 \text{GeV}) & 0.0336 & -3.119 & 0.0968 &-2.5614 & 0.1113
&-2.434 &  0.1163
&-2.389 \\
A_{FB} (M_{t\bar{t}}<450 \text{GeV}) & 0.0453 &-1.616 & 0.126 &-1.3196 & 0.1446
&-1.253 & 0.151
&-1.23 \\
A_{FB} (|\Delta Y|>1) & 0.0154 & 1.078 & 0.0282 & 1.1585 & 0.0285
&1.1602 &
0.0284 &1.159\\
A_{FB} (|\Delta Y|<1) & 0.0161 & 0.234 & 0.0358 &0.3932 & 0.038 &0.414 &
0.0391 &
0.420\\
\hline
\hline
\chi^2 & & 14.83 & & 11.24 & & 10.48 & & 10.22 \\
\hline
\chi^2/\nu & & 3.71 & & 2.81 & & 2.62 & & 2.55 \\
\hline
|f^u_{13}| & & 0.549 & & 1.319 & & 2.105 & & 2.905 \\
\hline
\end{array}
\end{math}
\end{small}
\caption{Results of $\chi^2$ analysis carried out for different values of $m_{\sgm_3}$. The best
fitted values for each observables along with their respective pulls are shown. The pull measures
the deviation in the fitted value of the observable from its mean value. For NP 
contributions, the number of d.o.f. is 4 (No. of observables $-$ No. of parameters.)}
\label{tab:sig3}
\end{table}

The results of the $\chi^2$ analysis are shown in Table \ref{tab:sig3}. We show the best-fit
values of all the observables along with their respective pulls. The minimum values of $\chi^2$ and 
the corresponding values of parameter $f^u_{13}$ obtained at the minimum are shown for different
masses of \St. The overall fits get better with increase in $m_{\sgm_3}$. For all masses, we get the largest pulls
corresponding to $A_{FB}$ in the $M_{t\bar{t}}>450$ GeV region where it gives more than $2\sigma$
deviation. All the other observables can be fitted within $1.2\sigma$. Although the total $\chi^2$ for NP contribution is 
better than $\chi^2$ for the SM, the $\chi^2/\nu$ values are worse than the SM value for smaller values of 
\St~ mass. The $\chi^2/\nu$ shows slight improvement relative to the SM only for 
masses greater than 1.5 TeV. Hence, the sextet \St~ in our model can satisfy the total 
cross section and the total asymmetry within 1$\sigma$ while it is incompatible with asymmetries 
in the large invariant mass region and the large rapidity region for the same parameter space. \\

\begin{figure}
\begin{center}
\includegraphics[width=3.2in,angle=-90]{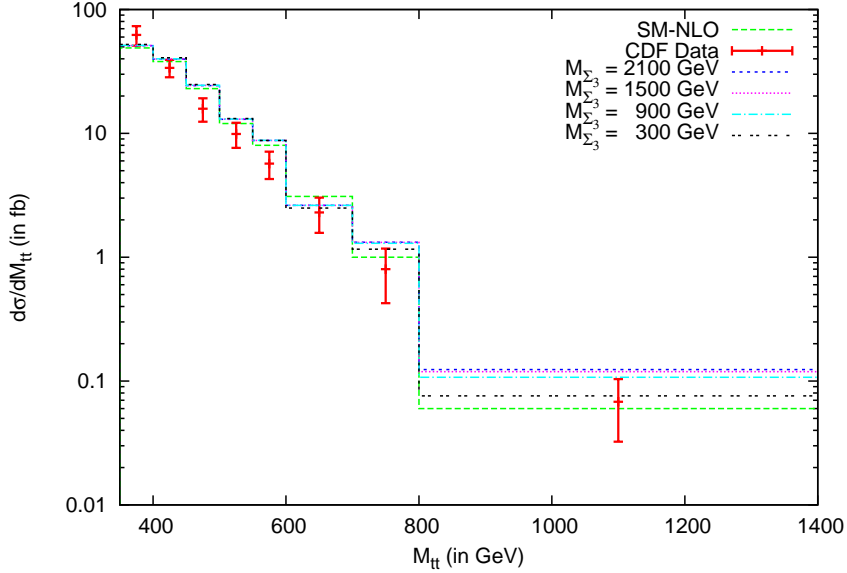}
\caption{ The \tbar invariant mass distribution for NLO-SM, and for various masses of colored
sextet \St ~for 
the best fitted values from our $\chi^2$ analysis.} 
\label{inv-mass-uu}
\end{center}
\end{figure}

An empirical relation between 
$f^u_{13}$ and $m_{\Sigma_3}$ is obtained and can be written in approximate form as :
\be\label{fit-m-g-u}
|f^u_{13}|=0.148 + 1.31\frac{m_{\Sigma_3}}{\text{1 TeV}}.
\ee
  Another important constraint in the \tbar production comes from the invariant mass distribution of
\tbar pair. This distribution has been measured by CDF collaboration and is shown in Fig. \ref{inv-mass-uu} 
for various values of \St ~masses with CDF data and SM-NLO prediction. We use the best fit values of  
coupling $f_{13}^u$ for various masses as shown in Table \ref{tab:sig3} for evaluating the 
contribution of NP to the $M_{t\bar{t}}$ distribution. The SM contribution to the $d\sigma/dM_{t\bar{t}}$ distribution 
in Fig. \ref{inv-mass-uu}  has been evaluated to the full NLO order as given in Ref. 
\cite{Cao:2010zb}. 
While evaluating the contributions of NP and its interference with the SM to invariant $M_{t\bar{t}}$ 
distribution, we multiply the contribution with K-factor of 1.3. However, it is highly desirable to 
include full NLO corrections to NP to make more reliable prediction on the invariant mass distribution.
We see that lower values of \St ~masses fit the distribution better than larger values of \St ~masses.

\subsection{Diquark $(6,1,\frac{1}{3})$}

\begin{figure}
\begin{center}
\includegraphics[width=3.0in]{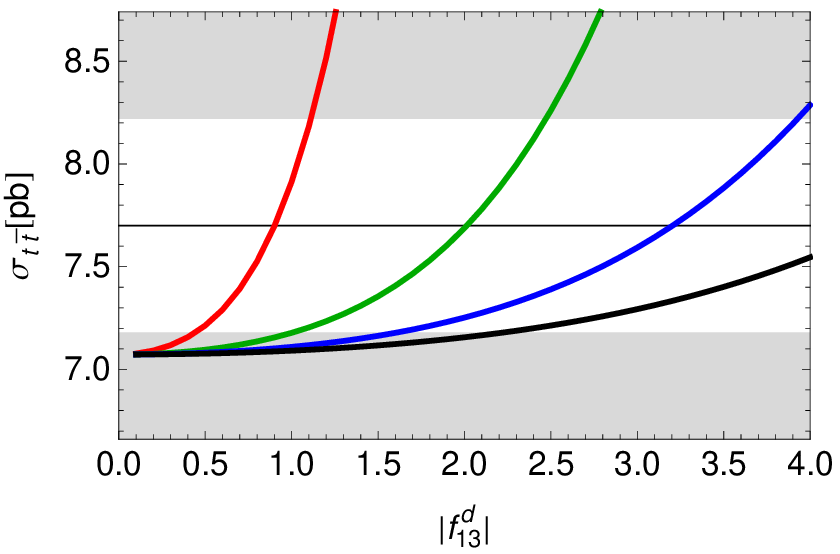}
\includegraphics[width=3.15in]{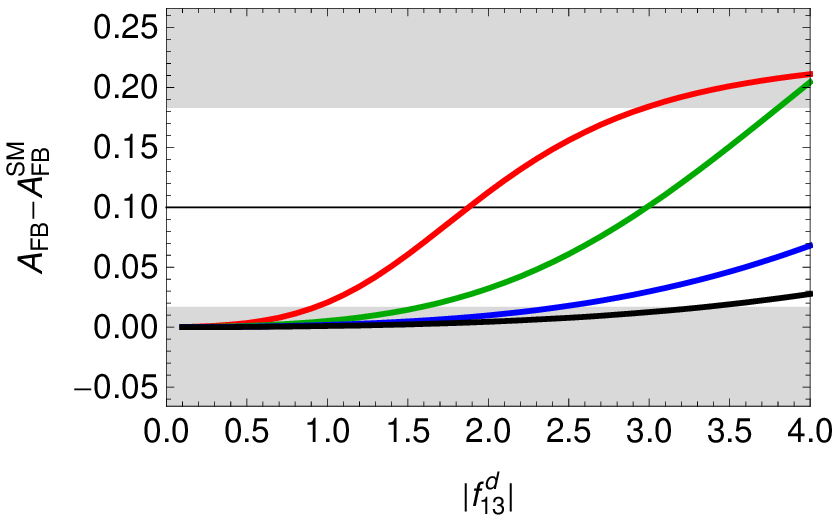} 
\caption{ The \tbar production cross section (left) and the forward backward asymmetry (right) 
as a function of the coupling $f^d_{13}$ for masses $m_{\Sigma_4}= $ 300 GeV (Red), 900 GeV (Green), 
1500 GeV (Blue) and 2100 GeV (Black). The unshaded and the shaded region correspond to $1\sigma$ 
and $2\sigma$ experimental bounds respectively.} 
\label{cs-afb-dd}
\end{center}
\end{figure}

Next, we study the diquark \Sf ~to look at its effect on \tbar pair production 
at the Tevatron. The contribution of \Sf ~to \tbar production has been shown in 
Fig. \ref{feyngraph} and proceeds through a \dbar initial state. In Fig. \ref{cs-afb-dd}, 
we plot the cross section and the forward backward asymmetry for \tbar production 
at Tevatron as a function of coupling $f^d_{13}$ for four different masses of colored sextet scalar
\Sf. As stated earlier, in showing the contribution from new physics, we subtract the SM contribution 
from the experimentally measured value of $\fba$. 

Because of the fact that the PDF for the d-quark is smaller than that for the u-quark, we need larger values of the 
coupling $f^d_{13}$ to generate the contribution to the observables of \tbar pair production. 
From Fig. \ref{cs-afb-dd}, just like for \St, \Sf ~of mass 300 GeV can barely satisfy both
$\sigma^{t\bar{t}}$ and $\fba$ for a very narrow range of coupling $f^d_{13}$ that too at 2$\sigma$
but with different range of coupling.
For larger masses \Sf ~can satisfy both the constraints for large range of 
coupling $f^d_{13}$ within 1$\sigma$ of experimental bound but with a wider and different 
range of couplings as compared to \St.
\begin{table} [h]
\begin{small}
\begin{math}
\begin{array}{|l|cc|cc|cc|cc|}
\hline
 &  \multicolumn{2}{c|} {m_{\Sigma_4}=300 \text{GeV}} &
\multicolumn{2}{c|} {m_{\Sigma_4}=900
\text{GeV}} & \multicolumn{2}{c|} {m_{\Sigma_4}=1.5 \text{TeV}} &
\multicolumn{2}{c|}
{m_{\Sigma_4}=2.1 \text{TeV}} \\
\hline
 \text{Observables} & \text{Fit} & \text{pull} &
\text{Fit} &
\text{pull} & \text{Fit} & \text{pull} & \text{Fit} &
\text{pull} \\
\hline
\text{Cross section} & 8.0157 & 0.6071 & 8.2649 & 1.0863 &
8.2734 & 1.1027 & 8.2968 &
1.1478 \\
A_{FB} & 0.0230 &-0.9277 & 0.0604 &-0.4775 &
0.0490 &-0.6143 & 0.0710
&-0.3499 \\
A_{FB} (M_{t\bar{t}}>450 \text{GeV})& 0.0299 & -3.1515 & 0.0985 &-2.546 & 0.1041 & -2.4972 &
0.1192 & -2.3636 \\
A_{FB} (M_{t\bar{t}}<450 \text{GeV})& 0.0428 & -1.6249 & 0.1299 &-1.3070 & 0.1365 & -1.2827 &
0.1547 &-1.2164 \\
A_{FB}(|\Delta Y|>1) & 0.0184 &1.0971 & 0.0345 &1.201 &
0.0338 &1.1934
& 0.0350
& 1.2012 \\
A_{FB} (|\Delta Y|<1) & 0.0173 & 0.2441 & 0.0405 & 0.4313 & 0.0405 &
0.4313 &
0.0429 & 0.4513 \\
\hline
\hline
\chi^2&  & 15.07 & & 11.23 & & 11.09 & & 10.15  \\
\hline
\chi^2/\nu&  & 3.78  & &  2.81 & &  2.78 & & 2.54  \\
\hline
|f^d_{13}|&  &1.041 & &2.508 & &3.875 & &5.527  \\
\hline
\end{array}
\end{math}
\end{small}
\caption{Results of $\chi^2$ analysis carried out for different values of $m_{\sgm_4}$. The best
fitted values for each observables along with their respective pulls are shown. The pull measures
the deviation in the fitted value of the observable from its mean value. $\nu$ denote number of degree of freedom. For NP 
contributions, the number of degree of freedom is 4 (No. of observables $-$ No. of parameters.}
\label{tab:sig4}
\end{table}

The results of the $\chi^2$ analysis for \Sf~ are shown in Table \ref{tab:sig4}. 
Similar to the previous case, the overall fits get better with increase in $m_{\sgm_4}$. 
As seen earlier, the $\chi^2/\nu$ values are worse than the SM value for smaller values of 
\Sf~ mass. Similar to the previous case, the best-fit relation between $f^d_{13}$ and $m_{\Sigma_4}$ can be
put into approximate form as :
\be\label{fit-m-g-d}
|f^d_{13}|=0.273 + 2.48\frac{m_{\Sigma_4}}{\text{1 TeV}}.
\ee
The perturbativity argument regarding the strength of a generic coupling `$g$' requires $g^2/4\pi<1$
, which allows, in principle, the coupling `$g$' to be as large as $\sim3.5$. 
In Figs. \ref{cs-afb-uu} and \ref{cs-afb-dd}, we show the cross section and the $\fba$ up to 
$f^{u,d}_{13}=4$. 
However, these are not the canonical couplings which enter in the Lagrangian of Eq. \ref{intterm}.
$f_{13}^{u,d}$ 
is related to the canonical coupling $F_{13}$ according to the Eq. \ref{coupling}. Using this
relation, 
we find that the values of the couplings $f_{13}^{u,d}$ which we use in our analysis satisfy the
perturbativity.

\begin{figure}
\begin{center}
\includegraphics[width=3.2in,angle=-90]{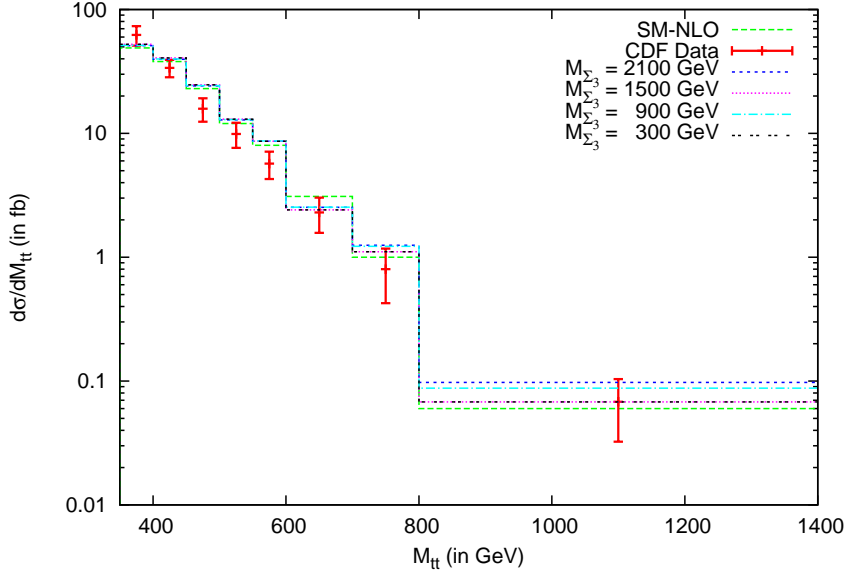}
\caption{ The \tbar invariant mass distribution for NLO-SM, and for various masses of colored
sextet \Sf~for the best fitted values from our $\chi^2$ analysis.} 
\label{inv-mass-dd}
\end{center}
\end{figure}

The invariant mass distribution of \tbar pair corresponding to contribution of \Sf~ is shown 
in Fig. \ref{inv-mass-dd} for various values of \St ~masses with CDF data 
and SM-NLO prediction. We use the best fit values of  
coupling $f_{13}^d$ for various masses as shown in Table \ref{tab:sig4} for evaluating the 
contribution of NP to the $M_{t\bar{t}}$ distribution. From the fig., we see that all values of \St ~masses 
are more compatible with the distribution in the large $M_{t\bar{t}}$ bin and fit the 
distribution better than \St. However, there is a little tension in the distribution for bins 
450 GeV-500 GeV and 550 GeV-600 GeV.

We now discuss the constraints on masses and couplings of the colored sextet \St~ and \Sf. These
constraints have 
been discussed in detail in Ref. \cite{Arnold:2009ay} where authors have analyzed the electroweak
precision data (EWPD) to obtain 
lower bound on the mass of the sextet scalars. 
They find that EWPD does not give a lower bound much above 100 GeV. The 
constraint is weak because there is no custodial $SU(2)_c$ violation. The most robust bound on sextet masses comes from 
direct search of these scalars at LEP-II putting a lower limit of 105 GeV on their masses. At Tevatron, the most stringent 
bound comes \cite{Aaltonen:2008dn} from the search of narrow resonances in the dijet mass spectrum. They reported lower mass bound for 
diquark to be $290$ GeV. The \St ~can produce same sign dileptons through decay into two top quarks 
while \Sf ~safely avoids same sign dilepton constraints. The constraints on the mass of \St ~from
the search of same sign 
dilepton signature is however weaker than the bound which comes from the search of narrow
resonances. 

The other stringent constraints come from low energy processes such as $D^0-\bar{D}^0$ mixing. 
The contributions of the sextet colored scalars to this mixing has been studied in some detail 
in Ref. \cite{Chen:2009xjb}. \St ~contributes in $D^0-\bar{D}^0$ mixing at tree level while \Sf 
~contributes 
through a box diagram. The bound on parameter for the \St ~is $|Re(f^u_{22}f^{u*}_{11})|\lesssim
5.76\times 10^{-7}$ 
for $m_H=1$ TeV. 
The bound on coupling of \Sf ~is $|f^{d*}_{12}f^d_{11}|^2\lesssim1.7\times 10^{-10}$  for $m_H=1$
TeV. 
However, these bounds can be relaxed if the couplings to second generation is minimized which 
we do in our analysis for search these scalars at LHC.

\section{Sextet diquarks \St ~and \Sf ~at LHC}

The detailed phenomenology of sextet diquarks has already been performed in 
Refs. \cite{Mohapatra:2007af,Han:2009ya,Chen:2008hh} where they found that such scalars can be
discovered at the LHC 
with masses around few GeV to 2 TeV. We have already shown in previous sections that \St ~and \Sf
~can have masses 
around this mass range in order to achieve unification at the GUT scale and explain the anomaly in
the \tbar 
production forward backward asymmetry at Tevatron.

The colored sextet scalar diquarks can be produced in the following channels at LHC:
\begin{enumerate}
 \item Resonant production in s-channel : pp$\rightarrow\Sigma_{3}/\Sigma_{4}\rightarrow t+u/d+X$,
 \item Pair-production : pp$\rightarrow\Sigma_{3}\Sigma_{3}^*/\Sigma_{4}\Sigma_{4}^*+X\rightarrow
(t+u/d)+(\bar{t}+\bar{u}/\bar{d})+X$,
 \item Single production with top quarks :
pp$\rightarrow\Sigma_{3,4}\bar{t}/\Sigma_{3,4}^*t+X$.
\end{enumerate}

The channel 1 has been explored in detail in Refs. \cite{Mohapatra:2007af,Han:2009ya}. 
The matrix amplitude squared and the resulting parton level cross section have been given 
in these references. The full NLO QCD corrections to \qbar annihilation to scalar sextet 
diquark resonant state has been performed in Ref. \cite{Han:2009ya}. 
In this channel, uu and ud initial states dominate over 
$\bar{u}\bar{u}$ and $\bar{u}\bar{d}$ initial states because of the large PDFs 
of quarks compared to antiquarks. Also, it is interesting to compare the cross sections 
corresponding to \St ~and \Sf ~production. The \St ~and \Sf ~get contributions from uu and ud
initial state 
respectively. The \St ~gets enhancement due to large u-quark PDFs while \Sf ~gets enhancement 
from two sources : (a) due to combinatorics from initial state, 
the luminosity of \Sf ~is $du\otimes ud$ while that of \St ~is $uu$, and (b) from the 
relations \ref{fit-m-g-u} and \ref{fit-m-g-d}, it can be seen that the coupling $f_{13}^d$ 
is larger than $f_{13}^u$. Hence, the cross section for \Sf ~is almost 5 times larger than 
\St ~production cross section for low masses and is about 1.5 times larger for large values 
of $\Sigma$'s masses. The best strategy to discover $\Sigma_{3,4}$ in this channel would 
be to determine the invariant mass distribution of $\bar{t}+$j and look for narrow resonances 
of $\Sigma_{3,4}$ as discussed in detail in 
Ref. \cite{Mohapatra:2007af}. 

\begin{figure}[h]
\begin{center}
\includegraphics[width=2.2in,angle=-90]{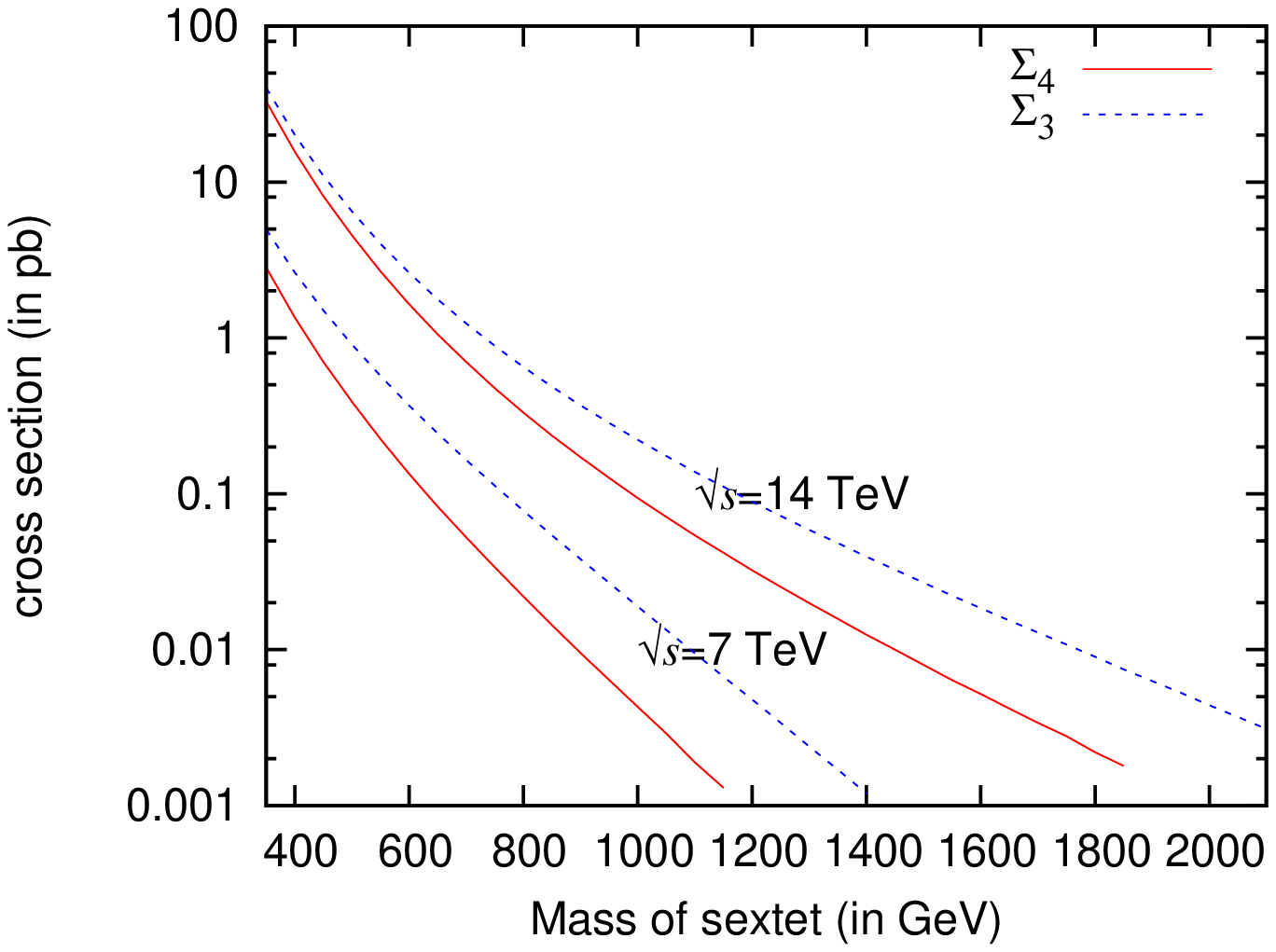}
\includegraphics[width=2.2in,angle=-90]{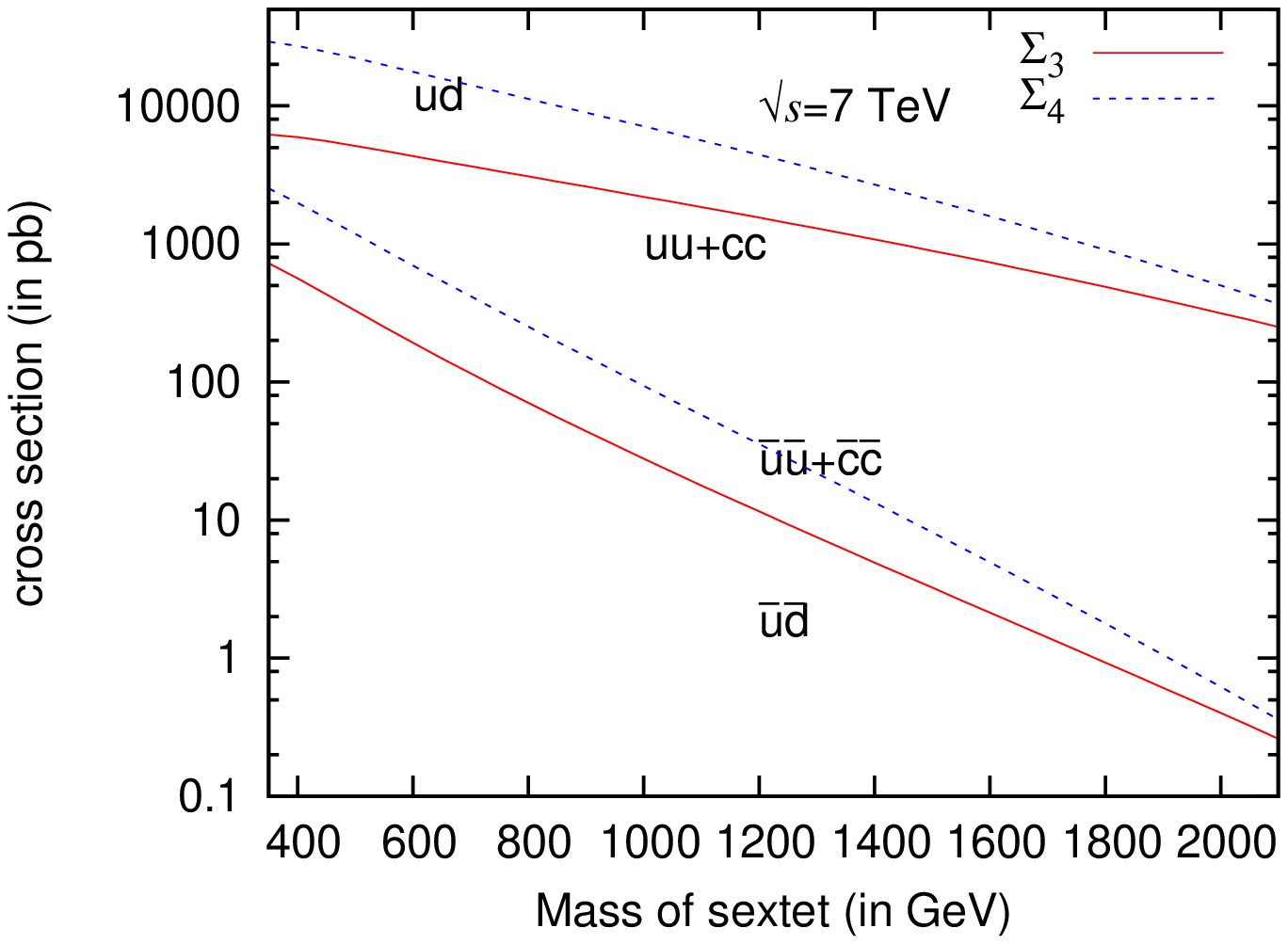} 
\caption{ The cross section for pair production $pp\rightarrow \Sigma_{3,4}\Sigma_{3,4}^*+X$ (left) 
and s-channel resonant production (right) of $pp\rightarrow \Sigma_{3,4}\rightarrow t+u/d+X$ 
with various initial states at the LHC. The values of the couplings $f^u_{13}$ and $f^d_{13}$
are evaluated using relation \ref{fit-m-g-u} and \ref{fit-m-g-d} respectively. 
We assume couplings $f^{d,u}_{13}=f^{d,u}_{11}$.} 
\label{cs-LHC1}
\end{center}
\end{figure}

The channel 2 has been explored in detail in Ref. \cite{Chen:2008hh}. The production process is 
mediated through QCD interactions through gg fusion and \qbar annihilation and 
hence depends only on sextet masses. The matrix amplitude squared and the resulting parton 
level cross section have been given in this reference. The channel in which \St ~decays to  
$tt\bar{t}\bar{t}$ has been analyzed in great detail in Ref. \cite{Chen:2008hh} for 14 TeV LHC. They
propose a 
reconstruction in the multijet plus same-sign dilepton with missing transverse energy samples to search 
for $tt\bar{t}\bar{t}$ final states from sextet scalar production. The decays of \Sf ~would yield 
$(t+j)+(\bar{t}+j)$ which can be probed in 8-j channel of which two are b-jets and all jets are 
hard jets. The cross sections for both \St ~and \Sf ~are large enough so that they can be 
discovered in lower mass range at 14 TeV LHC. In Fig. \ref{cs-LHC1}, we show production 
cross sections for channel 1 and 2 for various cm energies of LHC and for various possible initial states.
To calculate cross sections, we evaluate couplings $f_{13}^u$ and $f_{13}^d$ from relations 
\ref{fit-m-g-u} and \ref{fit-m-g-d} respectively and assume $f^u_{13}=f^u_{11}$ and $f^d_{13}=f^d_{11}$.

\begin{figure}[h]
\begin{center}
\includegraphics[width=3.2in,angle=-90]{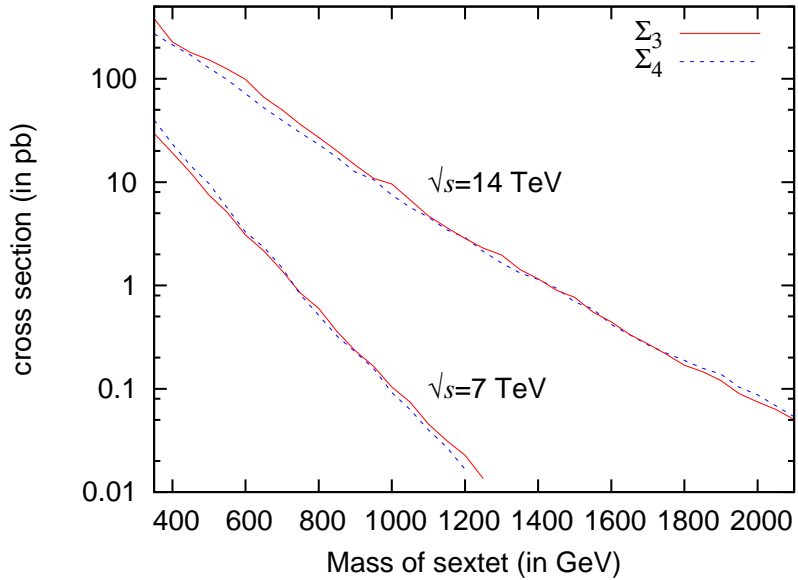}
\caption{ The cross section for process $pp\rightarrow \Sigma_{3,4}\bar{t}+X$ 
at the LHC for two cm of energies. The values of the couplings $f^u_{13}$ and $f^d_{13}$
are evaluated using relation \ref{fit-m-g-u} and \ref{fit-m-g-d} respectively. 
We assume couplings $f^{d,u}_{13}=f^{d,u}_{11}$.} 
\label{cs-LHC2}
\end{center}
\end{figure}

The \St~and \Sf~sextets can also be produced in association with antitop quarks. 
The cross sections for production of \St$+\bar{t}$ and \Sf$+\bar{t}$ pair have 
been shown in Fig. \ref{cs-LHC2}. 
Because of the large couplings of \Sf, the cross section for
\Sf ~production is comparable to that for \St~production. We see that for the interesting mass 
range of the sextets, the cross section is of the order of SM \tbar cross section which makes 
this channel very promising. The search strategy in this channel would be to look for $t\bar{t}j$ 
final states and search for resonances in the invariant mass of light jet with antitop quark.

\section{Summary}

In this paper, we have investigated the role of colored scalars as a possible explanation of large
forward-backward asymmetry in \tbar pair production at Tevatron. We consider a particular
non-supersymmetric \10 model where such scalars reside in $\overline{126}$ dimensional scalar
representation which plays a crucial part in GUT symmetry breaking and also generates small neutrino
masses through seesaw mechanism. We find that either \St~or \Sf~colored sextet submultiplet of
$\overline{126}$ can remain light and provide viable gauge coupling unification consistent with
the present bounds on proton decay.

Colored scalars in the context of forward-backward asymmetry at Tevatron have been studied in 
Refs. \cite{arhrib,tait,dorsner1}. All these papers are based on old CDF data and do not include new
observations. In this paper, we show that the contributions of light colored sextet scalars \St~and \Sf~of mass $>300$ GeV can enhance the 
forward-backward asymmetry without spoiling the $\sigma(t\bar{t})$ and the invariant mass distribution. We perform a 
$\chi^2$ analysis to simultaneously fit all the new observables along with total asymmetry and total cross section and find the best fitted 
values of the coupling for various masses. We find that all the observables can be fitted within $1.3\sigma$ of experiment errors 
except for the $\fba$ in $M_{t\bar{t}}>450$ GeV region where we can fit the observation at $2.3\sigma$. From the $\chi^2$ analysis, we 
conclude that the colored sextet scalars of masses 1.5 TeV-2.1 TeV can provide a marginal improvement over the SM observations if all observations are simultaneously 
considered in the fit.

In this paper, we focus on the study of light colored sextet scalars in \10 model and their effects to the $\fba$ 
in the light of new observations reported by CDF. We have shown that such scalars can emerge in particular
\10 model having masses of the order of TeV scale. However the detailed analysis carried
out in context of new and old CDF data are applicable to any sextet scalar.

We also discuss the various production mechanisms of \St~and \Sf~at the LHC and find that these 
scalars will have observable cross section to be discovered in future. These scalars can be produced 
in pairs owing to pure QCD interactions and this channels is promising at LHC rather than Tevatron. They 
can also be produced in s-channel resonance and then can be probed in their decay to $t+j$ events. Also, the other 
promising channel is to search them in single production in association with anti top quarks. The best strategy 
to search for these scalars in all these channels would be to look for narrow resonance in invariant mass of 
top quarks with light jets.\\
 
\noindent{\bf Acknowledgements}\\
 We would like to thank Anjan S. Joshipura, Namit Mahajan and Saurabh D.
Rindani for many useful discussions and suggestions at the different stages of
this work and also for reading the manuscript. We also thank Sandip Pakvasa for
useful discussion.\\

\noindent{\it Note Added:} While we were finishing the present work, Ref.
 \cite{Grinstein} was appeared  dealing with same kind of colored sextet state
$(\bar{6},1,-\frac{4}{3})$. Our results for \St~ scalar qualitatively agree with
their results. Compared to that work, we also study another sextet state and
show that such scalars naturally emerge from theoretically well motivated \10
model.\\

\section{Appendix}
In this appendix, we list all the sub-multiplets of the 10($\phi$), $\overline{126}$($\Sigma$) and
45($\chi$) dimensional scalar representations of \10 and their contributions to $B_{ij}=B_i-B_j$
coefficients (where $B_i$, $i=1,2,3$ are one loop $\beta$ function coefficients for $U(1)_Y,
SU(2)_L$ and $SU(3)_c$ respectively). We also present these sub-multiplets in terms of their
Pati-Salam subgroup ($SU(4) \times SU(2)_L \times SU(2)_R$) notations. $(R-,R0,R+)$ represents
components of the field which is triplet under $SU(2)_R$. The indices of the doublet of $SU(2)_L
(SU(2)_R)$ are denoted by $\alpha, \beta=1,2 (\alpha', \beta'=1',2')$. The index of the fundamental
4-plet of $SU(4)$ is denoted by $\mu(\nu)=\mubar(\nubar),4$ where $\mubar,\nubar=1,2,3$ represents
the $SU(3)$ subgroup indices. 

\begin{table}[ht]
\begin{center}
\begin{tabular}{|c|c|c|c|}
\hline
~Fields $(SU(3), SU(2), Y)$~ & ~Pati-Salam Notations~ & ~$\Delta B_{23}$~ & ~$\Delta B_{12}$~\\
\hline
\hline
$\phi_1 (3,1,-\frac{1}{3})$,~$\phibar_1 (\bar{3},1,\frac{1}{3}) $ 	&
$\phi_{\mubar4}$,~$\phi^{\mubar4}$  		&   ~$-\frac{1}{6}$~	& ~$\frac{1}{15}$~ \\
$\phi_2 (1,2,\frac{1}{2})$,~$\phibar_2 (1,2,-\frac{1}{2}) $ 	&
$\phi_{\alpha 1'}$,~$\phi^{\alpha}_{2'}$   		&   ~$\frac{1}{6}$~	&
~$-\frac{1}{15}$~ \\
\hline
$\chi_1 (8,1,0)$  	& $\chi_{\mubar}^{~\nubar}$		&   ~$-\frac{1}{2}$~	&  0\\
$\chi_2 (3,1,\frac{1}{3})$,~$\chibar_2 (\bar{3},1,-\frac{1}{3}) $ 	&
$\chi_{\mubar}^{~4}$,~$\chi_{4}^{~\nubar}$	&   ~$-\frac{1}{6}$~	& 
~$\frac{1}{15}$~\\
$\chi_{3,8} (1,1,0)$  	& $\chi^{(15)}$,~$\chi^{(R0)}$ 		&   0	&  0\\
$\chi_4 (3,2,-\frac{5}{6})$,~$\chibar_4 (\bar{3},2,\frac{5}{6}) $ 	&
$\chi_{\mubar 4 \alpha 2'}$,~$\chi_{\mubar \nubar \alpha 1'}$  		&   ~$\frac{1}{6}$~	&
~$\frac{1}{3}$~ \\
$\chi_5 (3,2,\frac{1}{6})$,~$\chibar_5 (\bar{3},2,-\frac{1}{6}) $ 	&
$\chi_{\mubar 4 \alpha 1'}$,~$\chi_{\mubar \nubar \alpha 2'}$  		&   ~$\frac{1}{6}$~	& 
~$-\frac{7}{15}$~\\
$\chi_6 (1,3,0)$  	& 	$\chi_{\alpha \beta }$	&  
~$\frac{1}{3}$~	&  ~$-\frac{1}{3}$~\\
$\chi_7 (1,1,1)$,~$\chibar_7 (1,1,-1) $ 	&
 $\chi^{(R+)}$,~$\chi^{(R-)}$ 		&   0	& ~$\frac{1}{5}$~ \\
\hline
$\sgm_1 (3,1,-\frac{1}{3})$,~$\sgmbar_1 (\bar{3},1,\frac{1}{3}) $ 	&
 $\sgm_{\mubar 4}$,~$\sgm^{\mubar4}$ 		&   ~$-\frac{1}{6}$~	& ~$\frac{1}{15}$~
\\
$\sgm_2 (1,2,\frac{1}{2})$,~$\sgmbar_2 (1,2,-\frac{1}{2}) $ 	&
$\sgm_{\alpha 1'}$,~$\sgm^{\alpha}_{2'}$  	&   ~$\frac{1}{6}$~	& ~$-\frac{1}{15}$~
\\
$\sgm_3 (6,1,\frac{4}{3})$  	& $\sgm_{\mubar \nubar}^{(R+)}$		&   ~$-\frac{5}{6}$~	& 
~$\frac{32}{15}$~\\
$\sgm_4 (6,1,\frac{1}{3})$  	& $\sgm_{\mubar \nubar}^{(R0)}$		&   ~$-\frac{5}{6}$~	& 
~$\frac{2}{15}$~\\
$\sgm_5 (6,1,-\frac{2}{3})$  	& $\sgm_{\mubar \nubar}^{(R-)}$		&   ~$-\frac{5}{6}$~	& 
~$\frac{8}{15}$~\\
$\sgm_6 (3,1,\frac{2}{3})$  	& $\sgm_{\mubar 4}^{(R+)}$		&   ~$-\frac{1}{6}$~	& 
~$\frac{4}{15}$~\\
$\sgm_7 (3,1,-\frac{1}{3})$  	& $\sgm_{\mubar 4}^{(R0)}$		&    ~$-\frac{1}{6}$~	&
~$\frac{1}{15}$~ \\
$\sgm_8 (3,1,-\frac{4}{3})$  	& $\sgm_{\mubar 4}^{(R-)}$		&    ~$-\frac{1}{6}$~	& 
~$\frac{1}{15}$~\\
$\sgm_9 (1,1,0)$  	& 	$\sgm_{4 4}^{(R+)}$	&   0	&  0\\
$\sgm_{10} (1,1,-1)$  	& 	$\sgm_{4 4}^{(R0)}$	&   0	&  ~$\frac{1}{15}$~\\
$\sgm_{11} (1,1,-2)$  	& 	$\sgm_{4 4}^{(R-)}$	&   0	&  ~$\frac{1}{15}$~\\
$\sgm_{12} (\bar{6},3,-\frac{1}{3})$  	&  $\sgm^{\mubar \nubar}_{~~\alpha \beta}$		&  
~$\frac{3}{2}$~	& ~$-\frac{18}{5}$~
\\
$\sgm_{13} (\bar{3},3,\frac{1}{3})$  	& $\sgm^{\mubar 4}_{~~\alpha \beta}$		&  
~$\frac{3}{2}$~ & ~$-\frac{9}{5}$~
\\
$\sgm_{14} (1,3,1)$  	& $\sgm^{4 4}_{~~\alpha \beta}$		&   ~$\frac{2}{3}$~
& ~$-\frac{1}{15}$~ \\
$\sgm_{15} (8,2,\frac{1}{2})$,~$\sgmbar_{15} (8,2,-\frac{1}{2}) $ 	&
 $\sgm_{\mubar~\alpha 1'}^{~\nubar }$,~$\sgm_{\mubar~\alpha 2'}^{~\nubar }$  		&  
~$-\frac{2}{3}$~	& ~$-\frac{8}{15}$~ \\
$\sgm_{16} (3,2,\frac{1}{6})$,~$\sgmbar_{16} (\bar{3},2,-\frac{1}{6}) $ 	&
$\sgm_{\mubar~\alpha 2'}^{~4 }$,~$\sgm_{4~\alpha 1'}^{~\nubar }$  		&  
~$\frac{1}{6}$~	&  ~$-\frac{7}{15}$~\\
$\sgm_{17} (3,2,\frac{7}{6})$,~$\sgmbar_{17} (\bar{3},2,-\frac{7}{6}) $ 	&
$\sgm_{\mubar~\alpha 1'}^{~4 }$,~$\sgm_{4~\alpha 2'}^{~\nubar }$  		&   ~$\frac{1}{6}$~
&  ~$\frac{17}{15}$~\\

\hline
\end{tabular}
\caption{Different sub-multiplets of 10($\phi$), $\overline{126}$($\Sigma$) and 45($\chi$)
dimensional scalar representations of \10 and their contribution to $B_{ij}$ coefficients of
Eq. \ref{beta}. Various notations used
are explained in text.}
\label{feilds}
\end{center}
\end{table}

\end{document}